\documentclass[mathleft]{an}

\usepackage{txfonts}
\usepackage{graphicx}
\usepackage{capt-of}
\usepackage{times}

\usepackage[authoryear]{natbib}
\bibpunct{(}{)}{;}{a}{}{,}

\defcitealias{ClarkeEnsslin2006}{CE06} 	
\defcitealias{Brentjens2008}{B08}

\usepackage{amsbsy}

\begin{document}

\title{Nonthermal emission from the radio relic of the galaxy cluster A2256}

\author{G. Siemieniec-Ozieblo\thanks{Corresponding author:
  \email{grazyna\@@oa.uj.edu.pl}}
\and B.M. Pasternak}
\titlerunning{Nonthermal emission from A2256}
\authorrunning{G. Siemieniec-Ozieblo \& B.M. Pasternak}
        
\institute{
Astronomical Observatory, Jagellonian University\\
Faculty of Mathematics, Physics and Computer Science\\
ul. Orla 171, 30--244 Krak\'ow, Poland}   

\received{XXXX}
\accepted{XXXX}
\publonline{XXXX}

\keywords{Galaxies: clusters: individual: A2256, radiation mechanisms: non-thermal, acceleration of particles.}

\abstract{
  We aim to obtain a consistent description of non-thermal emissions from Abell 2256 and to give a prediction for a gamma-ray emission from this galaxy cluster. Assuming that a radio relic illuminates a localization of an ongoing merger, and that both radio and non-thermal part of hard X-ray emission are due to electron component of cosmic rays filling the relic, we derived from radio and hard X-ray properties of the relic in A2256 the magnetic field strength and number densities for relativistic electrons and protons.
  Due to the interpretation of the radio relic as a structure formed just where a shock front is, we discuss a gamma-ray emission at the cluster periphery.
  The estimated strength of the magnetic field in the relic is equal to 0.05 $\mu$G, while the amplitude of the electron number density varies from $3 \cdot 10^{-4}$ to $3 \cdot 10^{-5}$ cm$^{-3}$ (respectively for the relic thickness of 50 to 500 kpc). We got a substantial degree of non-equipartition between cosmic rays and magnetic field in the relic region, where the CR pressure is approaching that of thermal gas.
  Our prediction for LOFAR is a synchrotron flux from the relic region of the order of $\sim$6 Jy at 60MHz and $\sim$10 Jy at 30MHz. The lower limit of the $\gamma$-ray flux from the relic region calculated for a hadronic channel is of the order of $10^{-12}$ erg cm$^{-2}$.
}
     
\maketitle

\section{Introduction}

Knowledge of a nonthermal particle content in galaxy clusters is significant for several reasons. We would like to understand what mechanisms are responsible for the acceleration of particles and what further processes occur to them to finally obtain the nonthermal emissions. Cosmic rays (CR) can thus be taken as adequate tracers of cosmic plasma physics. They also might contribute in a global energetic budget of intracluster medium (ICM) and could be dynamically important during cluster evolution.

In particular, diffuse nonthermal emission in radio band observed in several ($\sim$50) clusters \citep*[e.g.][]{GiovanniniTordiFeretti1999, KempnerSarazin2001}, as well as nonthermal fraction in X-ray flux suggested in a few of them (e.g. Coma, \citealt{FuscoFemianoOrlandini_etal2004}; A2256, \citealt*{FuscoFemianoLandiOrlandini2005}; A2199, \citealt{Kaastra_etal1999}; A2163, \citealt*{RephaeliGruberArieli2006}), contains a valuable information about presence of relativistic particles.  What is their origin and how it depends on localisation i.e. on a connection with a radio halo or a relic feature -- are the most important questions.
 
Do these diffuse radio structures are due to the primary or secondary component of cosmic ray electrons and how large might be the contribution of relativistic protons? Where we can see them and what flux is supposed to be generated?  If substantial -- can we detect a gamma-ray radiation from at least reach clusters? To answer these questions a consistent description of all potentially detectable nonthermal emissions is needed. Such a model including the broad band nonthermal emission reported both in the  radio and hard X-ray (HXR) regime should  be also confronted to the relevant characteristics of the thermal emission.

Modern radiotelescopes give relatively reach information about radio spectra and a polarization of a diffuse radio emission from galaxy clusters. The observational HXR emission status is uncomparably worse.
The search for nonthermal HXR emission with BeppoSAX/PDS resulted in a detection of excess above the thermal emission in seven clusters \citep{Nevalainen_etal2004}. For two of them: A2256 and A3667, the HXR excess was claimed to be localized in radio relics \citep{FuscoFemianoDalFiume_etal2000, FuscoFemianoDalFiume_etal2001}. The presence of HXR nonthermal excess has been also verified by INTEGRAL \citep[Ophiuchus cluster][]{Eckert_etal2008}. 

There are several ideas in the literature proposed for the origin of radio relics \citep[e.g.][]{EnsslinBiermann_etal1998, EnsslinGopalKrishna2001} and halos \citep[e.g.][]{Dennison1980, DolagEnsslin2000, Brunetti_etal2004, CassanoBrunetti2005}. They are founded on different acceleration mechanisms, which are based on the occurence of various magnetohydrodynamics shock waves and turbulences, with the diverse injection processes involved.
The dynamical state of a cluster is here of particular interest since in many cases subclusters are involved in an on-going merger -- producing a strong shock front. In this context a relic nature and its position is particularly important. Within the common interpretation of a radio relic as a shock wave tracer,  its radio properties provide the information about the shock parameters and the surrounding magnetized plasma. Here we focus our attention on the radio relic of Abell 2256 since relatively a lot  is known about this object.

In this paper it is assumed that a relic occurence is a symptom of an advanced stage of a merger process. Thus, the observed relic illuminates the localization of the current event which presents one of many such episodes in hierarchical history of galaxy cluster formation. While the previous mergers were responsible for the radio halo features, the recent one provides the new portion of matter/gas,  merger shock heating and nonthermal activity seen in the form of a relic. The simultaneous shock acceleration provides in the region of relic a new population of accelerated cosmic rays. For the case of A2256 we verify whether this freshly accelerated and injected electron population reproduces the observed relic radio features and the attributed to the relic hard X-ray excess (interpreted as inverse Compton scattering of cosmic microwave background photons) which might be produced by the same population, giving finally the consistent picture of the nonthermal relic medium. We also compare the polarization properties of the relic with the magnetic field strenght calculated at the cluster periphery as well as obtain the estimation of the gamma-ray flux from the relic volume.

\section{Simple relic model}

Understanding of processes responsible for a relic phenomenon is of a special importance because of its  localisation. Typically, its peripheral position at a border of a galaxy cluster makes it a "window" towards the larger structure -- cosmic-web filament. Therefore its thermal and nonthermal plasma diagnostics might give an insight into the warm-hot intergalactic medium (WHIM) and finally solve the missing barion problem in galaxy clusters. 

In the present paper we attempt  to show that all observed radio properties of the magnetized plasma in A2256 are consistent with the estimation of HXR flux. We thus compare the results of papers of \citet{ClarkeEnsslin2006} and \citet{Brentjens2008} (hereafter \citetalias{ClarkeEnsslin2006} and \citetalias{Brentjens2008}) on the one hand and those of \citet{FuscoFemianoLandiOrlandini2005}, on the other.
The goal of this multiband confrontation is an independent (of the method presented in \citetalias{Brentjens2008}) estimation of the relic magnetic field and the density of relativistic electrons there.

\begin{figure}
  \resizebox{\hsize}{!}{\includegraphics{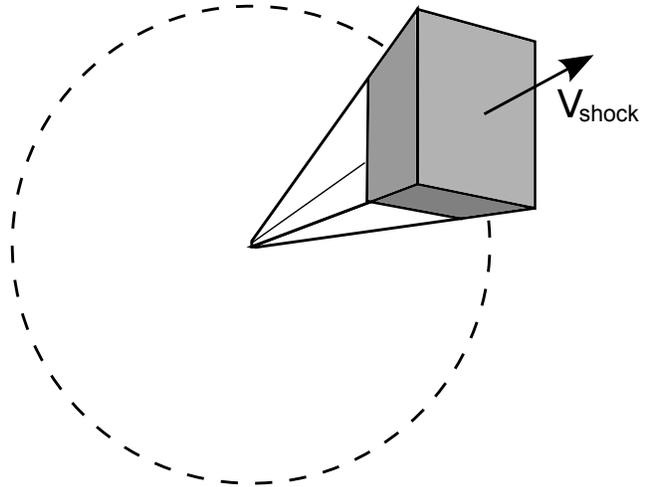}}
  \caption{Geometric representation of the relic in A2256.}
  \label{relic}
 \end{figure}

According to Clarke \& En\ss{}lin hypothesis the relic region covering an area of 1125$\times$520 kpc is settled on the front side of the cluster and reveals the merger localization. The main argument supporting this  thesis implies from the small value of the Faraday rotation measure dispersion of the relic (\citetalias{ClarkeEnsslin2006}). Assumig the relic  geometry shown in Fig. 11b in \citetalias{ClarkeEnsslin2006} and in Fig. 14 of \citetalias{Brentjens2008}, and the outward propagating merger shock, we present it as a rectangular layer observed at a small angle to the line of sight (Fig. \ref{relic}), i.e. propagating almost towards the observer.
  
Recently, in the literature, there were claimed several galaxy cluster relics that might be interpreted as outward propagating merger shocks \citep[e.g.][]{Bagchi_etal2006, vanWeeren_etal2009}. Some of them have been observed as spectacular double relic ring features \citep{Bonafede_etal2009}.
We accept the idea of an advanced merger shock propagating outward the cluster but leave unanswered the appealing question, whether sweeping by the same shock wave could be entirely responsible for the radio  halo phenomenon in A2256. 

The main radio properties of the cluster relic concerning the range of 350--1400 MHz are taken from \citetalias{ClarkeEnsslin2006} and \citetalias{Brentjens2008}. According to \citetalias{Brentjens2008}, the cluster flux can be modelled as a sum of two components: one that comes from the radio halo and the other one from the radio relic. The radio flux in the relic area (corrected for the halo flux) is 1.39$\pm$0.07 Jy ($\nu=351$ MHz) while the spectral index estimated in the above frequency range is taken as $\alpha_R=0.81$. We assume that the radio spectrum measured in this frequency range is a direct result of the emission from the shock-accelerated electrons. This young population has not steepened yet.

Statement of the polarized emission properties at both frequencies seems to be of a particular importance.
The high fractional polarization at 1400 MHz and lack of it at 350 MHz allows to estimate the Faraday depth of the relic environment which is in the following identified with the relic thickness $h$. The uniform Faraday rotation measure across the relic is consistent with the Galactic RM contribution and thus not attributed to the intracluster medium (\citetalias{ClarkeEnsslin2006}).

The hard X-ray detection made  by BeppoSAX/PDS \citep{FuscoFemianoLandiOrlandini2005} estimates the nonthermal flux at 20--80 keV band as $8.9\times10^{-12}$ erg cm$^{-2}$ s$^{-1}$ attributing this power law excess of X-ray emission to the radio relic located at the NW side of the cluster. A clear evidence of the nonthemal HXR emission in this cluster strongly supports the inverse Compton (IC) mechanism, due to the presence of the large region of freshly accelerated electrons. This property together with the multiple aspects of the radio emission from A2256 allows to test the physics of the relic,  i.e. the acceleration parameters as well as the merger shock features.

Assuming that the nonthermal X-ray emission originates in the radio relic and its power law index is close to 0.81 (within the fitted uncertainty range; $\alpha=1.5 _{-1.2}^{+0.3}$), the standard comparison of radio and nonthermal X-ray fluxes allows to derive both, the volume everaged relic magnetic field strength $B$ and the electron number density $N(\gamma) \mathrm{d}\gamma$, using \citet{RybickiLightman1979} formulae:
 \begin{eqnarray}
  F_s &=& \frac{N_0 K_1  a b h}{4\pi D^2} B^{\frac{p+1}{2}} \nu_r^{\frac{1-p}{2}}, \label{flux_r} \\
  F_{IC} &=& \frac{N_0 K_2  a b h}{4\pi D^2} \left ( k T \right ) ^{\frac{p+5}{2}} \epsilon_x^{\frac{1-p}{2}}. \label{flux_x}
 \end{eqnarray}
Here $\nu$ is the frequency of radio photons, $\epsilon$ is the X-ray photon energy and T -- the temperature of the cosmic microwave background (CMB) radiation. 

$ K_1$ and $ K_2$ are given by
 \begin{eqnarray}
  K_1 &=& \frac{\sqrt{3} q^3}{m c^2 (p+1)} \Gamma \! \left ( \frac{p}{4} + \frac{19}{12} \right ) \! \Gamma \! \left ( \frac{p}{4} - \frac{1}{12} \right ) \!\! \left ( \frac{2 \pi m c}{3q} \right ) ^{\frac{(1-p)}{2}} \\
  K_2 &=& \frac{8 \pi ^2 r_0^2}{h^3 c^2} J(p) \Gamma \! \left ( \frac{p+5}{2} \right ) \zeta \! \left ( \frac{p+5}{2} \right )
 \end{eqnarray}
where $J(p) = 2^{p+3} \frac{p^2 + 4p + 11}{(p+3)^2 (p+5) (p+1)}$ and the electron power law index $p$ is related to the radio spectral index $p = 1 + 2 \alpha$.

We assume the relativistic electrons fill isotropically the whole relic volume $a \cdot b \cdot h$ and their spectrum is given by
 \begin{equation}
  N(\gamma)\mathrm{d}\gamma = N_0 \gamma^{-p} \mathrm{d}\gamma, \qquad \gamma_{min}\leq\gamma\leq\gamma_{max}
 \end{equation}
where $h$ stands for an a priori unknown relic thickness, $\gamma$ is the electron  Lorentz factor and $N_0$ [cm$^{-3}$] is the electron distribution amplitude.

Following the argumentation in \citetalias{ClarkeEnsslin2006}, concerning the radio spectral trends in the relic, we assume the flat (0.81) spectrum does not change between 1400 - 351 MHz and below.

The strength of magnetic field $B$ can be directly derived from equations (\ref{flux_r}) and (\ref{flux_x}) and its value is equal to $0.05\pm 0.01\ \mu$G, while the amplitude of the electron number density  $N_0$, varies from $(3\pm3) \times 10^{-4}$  to $(3\pm3) \times 10^{-5}$cm$^{-3}$, respectively for $h$ from 50 to 500 kpc. Large uncertainties of $N_0$ are due to high sensivity of this parameter to even small changes in the spectral index $p$.

The above value of the strength of magnetic field $B$ is quite small although it falls well into the range $B=0.02-2 \mu$G estimated from depolarization properties of the relic in B08. CE06 found $B=1.5_{-0.6}^{+0.9}\mu$G and $B=3.3_{-1.2}^{+2.0}\mu$G using respectively classical and hadronic minimum energy condition, for the spectral index $\alpha=1.25$. Minimum energy condition is based on the assumption that nonthermal components of ICM are in an equilibrium state. This radio plasma pressure equilibrium seems to be implausible until the equipartition between the magnetic field and relativistic particles is achieved. One may suppose that it should be somehow reflected in $\mathcal{K}$ factor (ratio of energy density of CR protons to that of electrons) -- on the one hand, and the magnetic field filling factor -- on the other. Synchrotron filamentarity of the relic would also suggest that its $B$ filling factor is quite small and consequently indicate the state far distant from equipartition between the total energy densities of cosmic rays and that of the magnetic field. Thus if the hypothesis about relic being young formation created right behind a shock wave is true, this assumption might not be valid, explaining the much bigger values of the strength of magnetic field strength in CE06 than the value derived here.

Using the value of the cluster contribution to the Faraday rotation meassure RM$=-11\pm2$ rad m$^{-2}$ from B08 and the value of the mean magnetic field along the line of sight equal to the obtained value of $B=0.05\pm 0.01\ \mu$G one can estimate the thermal electron density in the relic. Assuming the relic thickness $h=50$ kpc and a coherence scale of the magnetic field $\geq h$, we get a rough value of $n_e\sim 5 \times 10^{-3}$cm$^{-3}$, quite similar to gas density at the west--central part of A2256 cluster $\sim3.6\times 10^{-3}$cm$^{-3}$ estimated from X--ray observations \citep{Sun_etal2002}.

We chose $h \in$ 100--200 kpc as a viable thickness range, comparing the derived above magnetic field strength with that obtained from the relic depolarization properties in \citetalias{Brentjens2008}.

Assuming the radio relic index as 0.81 and the corresponding electron spectrum $p\simeq2.6$, one gets for the merger shock compression $r=2.9$. Identification of the Faraday depth with the 100 kpc width of the Faraday rotating relic plasma matches the line of sight component of the magnetic field comparable to the above $B$ value (see Fig. 16 in \citetalias{Brentjens2008} for $r=2.9$).

Approximatively the same value of the volume averaged magnetic field strength and the one along the line of sight makes these different methods of $B$ estimation more reliably and consistent with its expected peripheral decreased value. It would also suggest a high $B$-field ordering in the relic. The width of the magnetized medium $h\sim$100 kpc seems also to be consistent with the filamentary structure seen at 1400 MHz.

The accelerated electrons radiating the synchrotron emission in the frequency range 350--1400 MHz have the Lorentz factor $\in (2.7 - 6)\times 10^4$, while those which upscatter the CMB photons within 20--80 keV band in a magnetic field $B\sim$0.05 $\mu$G place in (0.5 -- 0.9)$\times 10^4\  \gamma$-range. Thus, the latter should produce the synchrotron emission at $\nu \in (20 - 100 )$ MHz with the supposed synchrotron flux of the order of $\sim$6 Jy at 60 MHz and $\sim$10 Jy at 30 MHz (the future LOFAR surveys frequencies), for the estimated magnetic field strength and provided the spectral index is preserved at the same level.

The temperature distribution of the relic shown by Chandra map \citep{Sun_etal2002} at the border of A2256 may also constrain the merger shock compression. The hot region ($\sim$10 keV) coincident with the northern part of the relic and the temperature characteristic of the nearby sharp edge ($\sim$4 keV), cf. Fig. 3 \citet{Sun_etal2002}, supports the idea that the merger shock heating process could have occured.

However, the temperature gradient across the NW relic region (where $T_2/T_1\sim2.5$)  gives for the compression ratio a quite consistent but slightly smaller value, $r\simeq2.6$,
 \begin{equation}
  \frac{1}{r} = \left [ 4 \left ( \frac{T_2}{T_1} -1 \right ) ^2 +\frac{T_2}{T_1} \right ] ^{1/2} - 2 \left ( \frac{T_2}{T_1} -1 \right ).
 \end{equation}

Therefore, taking all the above into account, we cannot exclude that the single electron population produced in a merger shock is responsible for both, the X-ray and the radio nonthermal emission, although its energy spectrum extends from $\gamma_{min}\sim 10^3$ up to $\gamma_{max}\sim 6 \cdot 10^4$. Similar values of $B$ -- deduced from Faraday depolarization and independently from IC -- make the above hypothesis meaningful.

\section{Limits on relic gamma-ray emission}

The efficient electron acceleration in the forward propagating shock, suggested for the relic, allows to assume that also relativistic protons are effectively accelerated and confined in the same shocked region. Up to now there is no direct observational evidence of CR protons in galaxy clusters. However, their non-negligible contribution to the relativistic plasma pressure may have an essential impact on the ICM dynamics. 

To confirm the presence of the hadronic component, its interaction with other ICM components must be detected, mostly expected in a form of a diffuse gamma-ray emission. There are several processes producing $\gamma$-ray radiation. Depending on energy range some of them dominate. Thus, analyzing together the relevant emissions due to the leptonic part of CR and the anticipated hadronic one, the energy range of  both CR components has to be agreed. It means that to compare meaningfully the accelerated CR fractions we have to consider the case, where the energy range of relativistic protons corresponds to the range of synchrotron emiting electrons and thus determines the relevant gamma-ray process dominating there. 

In our case we restrict to analyzing the process operating in 10 to 30 GeV range. There are a few assumptions in the backgroud of such reasoning.

First we assume that for a given energy the acceleration time for electrons is the same as for protons. 

Secondly, contrary to common expectations concerning the impact of hadronic CR fraction mainly in central regions of  galaxy clusters, we discuss here the gamma-ray emission at the cluster periphery, i.e. in the relic volume. The main reason for that is the relic interpretation as a forward shock. Although  we have to do with the general trend that plasma density decreases with radius, the swept-up ICM gas is also compressed by the forward shock. CR density as well as the shock compression may increase as it goes outward. For the energy range of gamma-ray photons $\geq 1 $GeV,  relevant for Fermi satellite, the dominant  emission comes due to pion decay gamma-rays. As shown by \citet{AndoNagai2008}, the IC processes off the primary and secondary electrons give substatantially less contribution to gamma ray flux  than $\pi^0 $ decay in this energy range. Therefore, requiring the consistency in the multifrequency diagnostics in GeV gamma, HXR and radio energy range, based on simultaneously accelerated  CR particles, one can constrain the amount of relativistic protons.

To include CR protons, we assume that their energy density is comparable or by a factor of a few larger than the energy density in electrons.
In order to calculate the proton number we use here the scaling relations \citep{BeckKrause2005} comparing either the energy density of CR protons to that of electrons ($\mathcal{K}$ factor) or the ratio of number densities of protons and electrons $K_0$, within given energy range.  Both factors depend in general on acceleration process, propagation and energy losses. However, being in the regime where to a first approximation losses can be neglected \footnote{It is based on the fact, that time scale for acceleration is shorter than loss and escape time scales.}, it is usually supposed that they do not depend on energy. We take rather conservative values for $\mathcal{K}$ = 1 and 10 which correspond to $ K_0 = 10^3$ or a few $ \times 10^2$  \citep[cf.][Fig. 2]{BeckKrause2005}. In fact, even if we assume in relevant energy range that spectra for both injected CR species are proportional, the ratios $K_0(E)$ and $\mathcal{K}(E)$ are energy dependent due to losses.

Therefore, the finally implied values of $\rho^{(p)}$ and $N^{(p)}$ will give lower limits for energy density and number density of protons, where
 \begin{eqnarray}
  N^{(p)} & = & N_0^{(p)} \gamma^{-p} \\
  \rho^{(p)} & = & N_0^{(p)} \int_{\gamma_{min}}^{\gamma_{max}} \left ( \gamma m_p c^2 \right ) \gamma^{-p} \mathrm{d}\gamma .
 \end{eqnarray}
In the energy range relevant to the observed synchrotron radiation, i.e. \mbox{10--30 GeV}, the calculated electron energy density is $\sim 10^{-13}$ erg cm$^{-3}$.
 For values of $\mathcal{K}$ postulated above, one can read off the normalization factor for proton number $ N_0^{(p)}$, relating first 
$ \rho^{(p)}$ and $ N^{(p)} $ and comparing to the electron energy density.

For energy of 10 GeV this gives finally:
 \begin{equation}
  \left ( \frac{N^{(p)}}{N^{(e)}} \right ) _{10 \mathrm{GeV}} = 1.3 \left\{ \begin{array}{l}
									     10^4 \qquad \mathrm{for}\ \mathcal{K} = 1, \\
									     10^3 \qquad \mathrm{for}\ \mathcal{K} = 10.
									    \end{array} \right.
 \end{equation}
We postulate restrictively that the proton to electron energy density ratio $\mathcal{K}$ is constant in the considered energy range and we take for further consideration a more conservative value of  $\mathcal{K}\simeq1$. This scaling is compatible with the shock compression ratio $r\simeq3$ \citep[cf. Fig. 2 in][]{BeckKrause2005} and corresponds to the ratio $K_0\sim10^3$, larger then the canonical value of 100.

These values of $\mathcal{K} \sim 1$  and $K_0 \sim 10 ^ 3$ for  r = 3, \citep[Fig. 2 in][]{BeckKrause2005}, that is, the characteristic values of CR injected in the shock, can be treated as lower limits of the realistic ones influenced by energy losses. Thus the value of the density of protons resulting from them is the lower limit value as well.

Taking the proton density in relic tuned by electron density, in a way described above, these cosmic ray protons  injected into the relic medium through the same shock wave acceleration are following the power law $N_0^{(p)} \gamma^{-p}$ with $p=2.6$. Their interaction with the relic gas produce pions, whose decay yields $\gamma$-ray emission.
\newline\newline
Adopting the gamma-ray volume emissivity formula from Dermer's model \citep{PfrommerEnsslin2004}
 \begin{eqnarray}
  q_{\gamma}(r,E_{\gamma}) & \simeq & \sigma_{pp} c \, n_N(r) \xi^{2-\alpha_{\gamma}} \frac{\tilde{n}_{CRp}(r)}{GeV} \frac{4}{3 \alpha_{\gamma}} \left ( \frac{m_{\pi^0} c^2}{Gev} \right ) ^{-\alpha_{\gamma}} \nonumber \\
  & & \times \left [ \left( \frac{2 E_{\gamma}}{m_{\pi^0} c^2} \right ) ^{\delta_{\gamma}} + \left( \frac{2 E_{\gamma}}{m_{\pi^0} c^2} \right ) ^{-\delta_{\gamma}} \right ] ^{-\alpha_{\gamma}/\delta_{\gamma}}
 \end{eqnarray}
one gets the number of $\gamma$-ray photons per volume per energy range.
\newline
Here $\delta_{\gamma} = 0.14 \alpha_{\gamma}^{-1.6} + 0.44$, $\sigma_{pp} = 32 \times \left ( 0.96 + e^{4.4 -2.4\alpha_{\gamma}} \right )$ mbarn, $m_{\pi^0} c^2 / 2 \simeq 67.5$ MeV, $\xi = 2$, $n_N(r) = n_e(r) / (1-0.5X_{He})$ ($X_He = 0.24$), $\tilde{n}_p(r)$ comes from the kinetic CRp energy density $\rho_p$ \citep{PfrommerEnsslin2004}.

Since a $\gamma$-ray spectrum repeats a parental proton spectrum, then $\alpha_{\gamma}=p=2.6$. For the relic medium we took a constant value of the thermal electron density $n_e=4.8\times10^{-4}$ cm$^{-3}$ according to \citetalias{Brentjens2008}. The normalization factor $\tilde{n}$ related to the $N_0^{(p)}$ amplitude can be expressed through the proton to electron number density ratio $K_0$. 
\newline\newline
The $\gamma$-ray flux produced by the interaction of CR protons with ambient relic nuclei is given by 
 \begin{equation}
  F = \frac{1}{4\pi D^2} \int \mathrm{d}V \int_{1 GeV}^{\infty} \mathrm{d} E q(r,E)
 \end{equation}
where $V$ is the relic volume, $D$ is the luminosity distance.

One can see from the above that the relic volume effectively radiates gamma-rays in GeV range and gives in the frame of hadronic scenario the average flux of the order of $ 10^{-12}$ erg cm$^{-2}$. The flux value is underestimated since we ignored the CRs losses, preserving the spectral proportionality. If the losses are included both ratios, $K_0(E)$ and $\mathcal{K}(E)$, encrease with energy thus enlarging the proton number and the following gamma-ray emission. Independently, another potential factor to enhance the gamma-ray flux is a thermal electron density larger than the one used above.

\section{Summary}

There have been several papers describing and simulating in detail nonthermal emissions from galaxy clusters. Here we have presented a particular example of a multifrequency approach allowing to justify the relic is a region where a forward moving shock achieves a cluster periphery. 
Such an approach is quite closely bound to the question: do we already see the WHIM region linked to the large-scale structure filament and illuminated by the synchrotron emission from the relic? 
The relic volume of A2256 shows the shock front acceleration region. Shock driving in the NW direction seems to effficiently accelerate particles. Assuming that electrons and protons fill the relic volume, one can estimate peripheral ICM observables. Given the flat spectral index radio component and independently the HXR emission, presumably coinciding with the radio relic, we have derived, according to the polarization properties of the relic, the magnetic field strength and number densities for relativistic electrons and protons as well as the depth of the relic.

It is believed that a magnetic field decreases with a cluster radius and gains at an outskirt, as shown here, a strength value of $\sim$0.05 - 0.1 $\mu$G (close to the magnetic strength value in the filaments). Contrary to that, an energy density of CR should increase when approaching a relic. This trend suggests breaking of the equipartition principle. Irrelevance of the energy equipartition in the case of moving shock disqualifies this assumption as a method of a magnetic field strength estimate. Although the method based on IC emission may be accepted only if HXR excess located at the relic is confirmed and there is no break in the relic radio spectrum at low frequencies. We also estimated the radio flux at $\nu \in (20 - 100 )$ MHz which is of the order of $\sim$6 Jy at 60 MHz and $\sim$10 Jy at 30 MHz.

Assuming conservatively that the energy densities in leptonic and hadronic components of CR are comparable, we obtain for $\mathcal{K}=1$ electron and proton energy densities $\rho_e\simeq\rho_p\sim10^{-13}$ erg cm$^{-3}$ while for $\mathcal{K}=10$: $\rho_p\sim10^{-12}$ erg cm$^{-3}$. The energy density of magnetic field $B$ is much smaller $\sim10^{-16}$ erg cm$^{-3}$, indicating the substantial degree of non-equipartition between CRs and magnetic field. Comparing to thermal energy density,  $\rho_{th} \sim$ 1 eV cm$^{-3}$ ($\simeq 1.6\times10^{-12}$ erg cm$^{-3}$ for T $\sim$ 4 keV) we obtain for acceleration efficiency $ \rho_{nth}/\rho_{th} \sim $ (0.5 - 0.05). This local value of $\rho_{p} \simeq \rho_{e}$ characterizing the peripheries of galaxy cluster indicates relatively large acceleration efficiency. Directly following from this is significant amount of nonthermal pressure relative to thermal:
$ P_{CR} \sim 10^{-13}$ erg cm$^{-3}$, while $ P_{th} \sim 10^{-12}$ erg cm$^{-3}$ is comparable to ram pressure of forward moving shock, where  $P^{shock}_{ram} \propto \rho v^2 \simeq 10^{-12}$ erg cm$ ^{-3}$ for typical shock velocity  $\sim$ 1000 km/s.

These numbers allow to predict the expected $\gamma$-ray flux at the relic location. In the energy range $\geq$ 1 GeV, the ''contamination" to gamma-ray emission caused by IC upscattered relic photons on relativistic electrons is negligible. 
Since the $\pi^0 $-decay $\gamma$-rays trace the product of the CR$_{p}$ density and the plasma gas (pushed by the forward shock) density in the relic region, therefore we should expect the volume of the $\gamma$-ray emission should coincide with the relic. The $\gamma$-ray flux calculated for a hadronic channel is of the order of $10^{-12}$ erg cm$^{-2}$ and places well within the verification range of the FERMI Gamma-ray telescope. Thus, potentially observable emission comparable with the FERMI sensitivity will also allow to trace merger shocks at the borders of clusters  tied to filaments within the cosmic-web.

To summarize: a gamma-ray observation from a relic together with radio and HXR  will provide a "pinpointed" constraint on the acceleration mechanism of CR in galaxy clusters.

\acknowledgements
B.M.P. acknowledges support from the grant 92/N-ASTROSIM/2008/0 from Polish Ministry of Science and Higher Education.

\bibliographystyle{an}
\bibliography{bibliography}

\begin{thebibliography}{26}
\expandafter\ifx\csname natexlab\endcsname\relax\def\natexlab#1{#1}\fi

\bibitem[{{Ando} \& {Nagai}(2008)}]{AndoNagai2008}
{Ando}, S. \& {Nagai}, D. 2008, \mnras, 385, 2243

\bibitem[{{Bagchi} {et~al.}(2006){Bagchi}, {Durret}, {Neto}, \&
  {Paul}}]{Bagchi_etal2006}
{Bagchi}, J., {Durret}, F., {Neto}, G.~B.~L., \& {Paul}, S. 2006, Science, 314,
  791

\bibitem[{{Beck} \& {Krause}(2005)}]{BeckKrause2005}
{Beck}, R. \& {Krause}, M. 2005, Astronomische Nachrichten, 326, 414

\bibitem[{{Bonafede} {et~al.}(2009){Bonafede}, {Giovannini}, {Feretti},
  {Govoni}, \& {Murgia}}]{Bonafede_etal2009}
{Bonafede}, A., {Giovannini}, G., {Feretti}, L., {Govoni}, F., \& {Murgia}, M.
  2009, \aap, 494, 429

\bibitem[{{Brentjens}(2008)}]{Brentjens2008}
{Brentjens}, M.~A. 2008, \aap, 489, 69

\bibitem[{{Brunetti} {et~al.}(2004){Brunetti}, {Blasi}, {Cassano}, \&
  {Gabici}}]{Brunetti_etal2004}
{Brunetti}, G., {Blasi}, P., {Cassano}, R., \& {Gabici}, S. 2004, \mnras, 350,
  1174

\bibitem[{{Cassano} \& {Brunetti}(2005)}]{CassanoBrunetti2005}
{Cassano}, R. \& {Brunetti}, G. 2005, \mnras, 357, 1313

\bibitem[{{Clarke} \& {En{\ss}lin}(2006)}]{ClarkeEnsslin2006}
{Clarke}, T.~E. \& {En{\ss}lin}, T.~A. 2006, \aj, 131, 2900

\bibitem[{{Dennison}(1980)}]{Dennison1980}
{Dennison}, B. 1980, \apjl, 239, L93

\bibitem[{{Dolag} \& {En{\ss}lin}(2000)}]{DolagEnsslin2000}
{Dolag}, K. \& {En{\ss}lin}, T.~A. 2000, \aap, 362, 151

\bibitem[{{Eckert} {et~al.}(2008){Eckert}, {Produit}, {Paltani}, {Neronov}, \&
  {Courvoisier}}]{Eckert_etal2008}
{Eckert}, D., {Produit}, N., {Paltani}, S., {Neronov}, A., \& {Courvoisier}, T.
  2008, \aap, 479, 27

\bibitem[{{En{\ss}lin} {et~al.}(1998){En{\ss}lin}, {Biermann}, {Klein}, \&
  {Kohle}}]{EnsslinBiermann_etal1998}
{En{\ss}lin}, T.~A., {Biermann}, P.~L., {Klein}, U., \& {Kohle}, S. 1998, \aap,
  332, 395

\bibitem[{{En{\ss}lin} \& {Gopal-Krishna}(2001)}]{EnsslinGopalKrishna2001}
{En{\ss}lin}, T.~A. \& {Gopal-Krishna}. 2001, \aap, 366, 26

\bibitem[{{Fusco-Femiano} {et~al.}(2000){Fusco-Femiano}, {Dal Fiume}, {De
  Grandi}, {Feretti}, {Giovannini}, {Grandi}, {Malizia}, {Matt}, \&
  {Molendi}}]{FuscoFemianoDalFiume_etal2000}
{Fusco-Femiano}, R., {Dal Fiume}, D., {De Grandi}, S., {et~al.} 2000, \apjl,
  534, L7

\bibitem[{{Fusco-Femiano} {et~al.}(2001){Fusco-Femiano}, {Dal Fiume},
  {Orlandini}, {Brunetti}, {Feretti}, \&
  {Giovannini}}]{FuscoFemianoDalFiume_etal2001}
{Fusco-Femiano}, R., {Dal Fiume}, D., {Orlandini}, M., {et~al.} 2001, \apjl,
  552, L97

\bibitem[{{Fusco-Femiano} {et~al.}(2005){Fusco-Femiano}, {Landi} \&
  {Orlandini}}]{FuscoFemianoLandiOrlandini2005}
{Fusco-Femiano}, R., {Landi}, R., \& {Orlandini}, M. 2005, \apjl, 624, L69

\bibitem[{{Fusco-Femiano} {et~al.}(2004){Fusco-Femiano}, {Orlandini},
  {Brunetti}, {Feretti}, {Giovannini}, {Grandi}, \&
  {Setti}}]{FuscoFemianoOrlandini_etal2004}
{Fusco-Femiano}, R., {Orlandini}, M., {Brunetti}, G., {et~al.} 2004, \apjl,
  602, L73

\bibitem[{{Giovannini} {et~al.}(1999){Giovannini}, {Tordi} \&
  {Feretti}}]{GiovanniniTordiFeretti1999}
{Giovannini}, G., {Tordi}, M., \& {Feretti}, L. 1999, New Astronomy, 4, 141

\bibitem[{{Kaastra} {et~al.}(1999){Kaastra}, {Lieu}, {Mittaz}, {Bleeker},
  {Mewe}, {Colafrancesco}, \& {Lockman}}]{Kaastra_etal1999}
{Kaastra}, J.~S., {Lieu}, R., {Mittaz}, J.~P.~D., {et~al.} 1999, \apjl, 519,
  L119

\bibitem[{{Kempner} \& {Sarazin}(2001)}]{KempnerSarazin2001}
{Kempner}, J.~C. \& {Sarazin}, C.~L. 2001, \apj, 548, 639

\bibitem[{{Nevalainen} {et~al.}(2004){Nevalainen}, {Oosterbroek}, {Bonamente},
  \& {Colafrancesco}}]{Nevalainen_etal2004}
{Nevalainen}, J., {Oosterbroek}, T., {Bonamente}, M., \& {Colafrancesco}, S.
  2004, \apj, 608, 166

\bibitem[{{Pfrommer} \& {En{\ss}lin}(2004)}]{PfrommerEnsslin2004}
{Pfrommer}, C. \& {En{\ss}lin}, T.~A. 2004, \aap, 426, 777

\bibitem[{{Rephaeli} {et~al.}(2006){Rephaeli}, {Gruber} \&
  {Arieli}}]{RephaeliGruberArieli2006}
{Rephaeli}, Y., {Gruber}, D., \& {Arieli}, Y. 2006, \apj, 649, 673

\bibitem[{{Rybicki} \& {Lightman}(1986)}]{RybickiLightman1979}
{Rybicki}, G.~B. \& {Lightman}, A.~P. 1986, {Radiative Processes in
  Astrophysics} ({Wiley-VCH})

\bibitem[{{Sun} {et~al.}(2002){Sun}, {Murray}, {Markevitch}, \&
  {Vikhlinin}}]{Sun_etal2002}
{Sun}, M., {Murray}, S.~S., {Markevitch}, M., \& {Vikhlinin}, A. 2002, \apj,
  565, 867

\bibitem[{{van Weeren} {et~al.}(2009){van Weeren}, {R{\"o}ttgering}, {Bagchi},
  {Raychaudhury}, {Intema}, {Miniati}, {En{\ss}lin}, {Markevitch}, \&
  {Erben}}]{vanWeeren_etal2009}
{van Weeren}, R.~J., {R{\"o}ttgering}, H.~J.~A., {Bagchi}, J., {et~al.} 2009,
  \aap, 506, 1083

\end{thebibliography}

\end{document}